\documentclass[onecolumn]{aa} 
\usepackage{graphicx}
\usepackage{color}
\usepackage{ulem}             

\usepackage{txfonts}
\newcommand{\tnr}[1]{{#1}}    
\newcommand{\tjc}[1]{{#1}}

\def \eps {\varepsilon}

\def\etal{{et al.}}

\newcommand{\blanc}[1]{}

\def\eps{{\varepsilon}}

\def\be{\begin{equation}}
\def\ee{\end{equation}}

\def\etal{{\it et~al. }}
\def\aap{{\it Astronomy \& Astophysics }}
\def\jgr{{\it Journal of Geophysical Research }}

\begin{document}

\title{Constraining Ceres' interior from its Rotational Motion}

\author{N. Rambaux \inst{1,2}, J. Castillo-Rogez \inst{3}, V. Dehant \inst{4}, \and P. Kuchynka \inst{2,3}
}

   \institute{Universit\'e Pierre et Marie Curie, UPMC - Paris 06
   \and 
   IMCCE, Observatoire de Paris, CNRS UMR 8028,\\
 77 Avenue Denfert-Rochereau, 75014 Paris, France\\
              Tel.: +33 [0]1 40 51 22 63\\
              Fax: +33 [0]1 40 51 20 58\\
              \email{Nicolas.Rambaux@imcce.fr} \\
              \email{kuchynka@imcce.fr} 
         \and
         Jet Propulsion Laboratory, Caltech, Pasadena, USA\\
         \email{julie.c.castillo@jpl.nasa.gov}
         \and
	Royal Observatory of Belgium, 3 Avenue Circulaire, B-1180 Brussels, Belgium  \\            
	\email{vdehant@oma.be}
             }

   \date{Received xx; accepted xx}

 
  \abstract
 {Ceres is the most massive body of the asteroid belt and contains about 25 wt.\% (weight percent) of water. Understanding its thermal evolution and assessing its current state are major goals of the \textit{Dawn} Mission. Constraints on internal structure can be inferred from various observations. Especially, detailed knowledge of the rotational motion can help constrain the mass distribution inside the body, which in turn can lead to information on its geophysical history.} 
   {We investigate the signature of the interior on the rotational motion of Ceres and discuss possible future measurements performed by the spacecraft \textit{Dawn} that will help to constrain Ceres' internal structure. }  
   {We compute the polar motion, precession-nutation, and length-of-day variations. We estimate the amplitudes of the rigid and non-rigid response for these various motions for models of Ceres interior constrained by recent shape data and surface properties. }
   {As a general result, the amplitudes of oscillations in the rotation appear to be small, and their determination from spaceborne techniques will be challenging.  For example, the amplitudes of the semi-annual and annual nutations are around $\sim 364$ and $\sim 140$ milli-arcseconds, and they show little variation within the parametric space of interior models envisioned for Ceres. This, combined with the very long-period of the precession motion, requires very precise measurements.
We also estimate the timescale for Ceres' orientation to relax to a generalized Cassini State, and we find that the tidal dissipation within that object was \tnr{probably} too small to drive any significant damping of its obliquity since formation. 
However, combining the shape and gravity observations by \textit{Dawn} offers the prospect to identify departures of non-hydrostaticity at the global and regional scale, which will be instrumental in constraining Ceres' past and current thermal state. We also discuss the existence of a possible Chandler mode in the rotational motion of Ceres, whose potential excitation by endogenic and/or exogenic processes may help detect the presence of liquid reservoirs within the asteroid.  }


\titlerunning{Ceres' rotational motion}
   \maketitle
%

\section{Introduction}

The largest asteroid, Ceres, will be rendez-vous'd by the {\it{Dawn}} spacecraft in 2015. During the nominal mission, the asteroid will be mapped with high-resolution imaging in the visual and infra-red wavelengths (VIR instrument), and its gravity field will be determined to the tenth degree and order (McCord et al. 2011). A major goal of the {\it{Dawn}} Mission is to constrain the internal structure of Ceres, and especially quantify the extent of differentiation, and thus of internal evolution, achieved by this dwarf planet (Russell \etal 2007; McCord \etal 2011). 

While the primary objective of high-resolution camera VIR is to map the surface composition and surface feature, an interesting contribution of these observations will be to characterize Ceres' rotational motion. Rotation properties have proved to bring important constraints on the interiors of planetary bodies, (e.g., Mathews \etal 2002; Koot \etal 2010; Williams \etal 2001; Margot \etal 2007; Dehant \etal 2009) because they depend on mass distribution and viscoelastic properties.
Indeed, the departures from a uniform rotation and changes in the orientation of a body are responses to an external forcing such as the gravitational force of another celestial body of the solar system or of the Sun. These responses depend on the structure and composition of the interior. In particular the possible presence of liquid layers inside a body and the elastic or inelastic properties of the solid parts drive the object's response to the external forcing. In particular when there are liquid layers within a body, such as Mercury or the icy satellites, the librational response of the solid part is amplified. A similar effect is expected for the nutation when considering rapid orientation changes such as for the Earth and Mars (resonance to the Free Core Nutation). The material composing Ceres and its physical properties thus determine the rotational behavior of Ceres. The observation of the response of Ceres to any forcing may thus provide information on its interior.

The purpose of this paper is to investigate the temporal variations of Ceres' rotation in order to identify potential observations to be performed by the \textit{Dawn} Mission. So far, observations of Ceres' surface have been obtained with the \textit{Hubble Space Telescope} ({\it{HST}}), adaptive optics (Keck Telescope), and occultations. However, it has proved difficult to track surface landmarks with enough accuracy to constrain the polar orientation and rotation of the asteroid (Thomas \etal 2005; Carry \etal 2008; Drummond and Christou 2008). Due to the non-spherical shape of Ceres related to the centrifugal potential, the gravitational potential of the Sun exerts a non-zero torque on Ceres dynamical figure. Consequently, Ceres' axis of figure is expected to exhibit a precessional motion around its normal to the orbital plane and the periodic part of the torque generates a nutational motion along the precessional cone. In addition, in the reference frame of Ceres, the spin axis describes a polar motion or wobble around the figure axis, and, finally, tidal deformations arising from the Sun induce perturbations in its rotational velocity, leading to length-of-day variations (l.o.d.).

The first part of this article describes the shape and interior structure parameters used for modeling Ceres. Then, we review the observed pole position and discuss briefly its secular evolution. In Section 4, we introduce the main equations used to determine the rotational parameters of the body. The geophysical and rotational models are then combined to compute the rotational motion of Ceres as a function of polar motion and precession-nutation (Section~\ref{sec:rigide}). Based on these results, we discuss the prospects for characterizing Ceres' rotation with the {\it{Dawn}} Mission, and the constraints these observations will provide on Ceres' interior (Section~\ref{sec:geocons}).

\section{The shape and interior structure of Ceres}\label{sec:shape}

Available constraints on Ceres' interior come from ground-based and space telescope observations
and best estimates on the density and mean radius are gathered and discussed in McCord and Sotin (2005) and Castillo-Rogez and McCord (2010) (Table~\ref{tab:param}). Key information on the interior structure comes from shape data, which led to constraints on the mean moment of inertia, assuming that the object is in hydrostatic equilibrium. 
Evidence for Ceres' shape hydrostaticity was first suggested by Millis \etal (1987) based on 13 ground-based occultation observations, who concluded that the asteroid is an oblate spheroid. This configuration was confirmed by Thomas \etal (2005) from a dataset of $\sim$380 images obtained with the {\it{HST}} over 80\% of Ceres' rotation period, as well as from other observation campaigns (e.g., Carry \etal 2008; Drummond and Christou 2008). While it is not possible to rule out the possibility that Ceres bears non-hydrostatic anomalies with amplitudes of the order of the current uncertainty on the shape data ($\sim$ 2 km), we take as a working hypothesis that the asteroid is in hydrostatic equilibrium. We will discuss the validity of this assumption in Section~\ref{sec:geocons}. 

\begin{figure*}[htbp]
\begin{center}
\includegraphics[height=15cm]{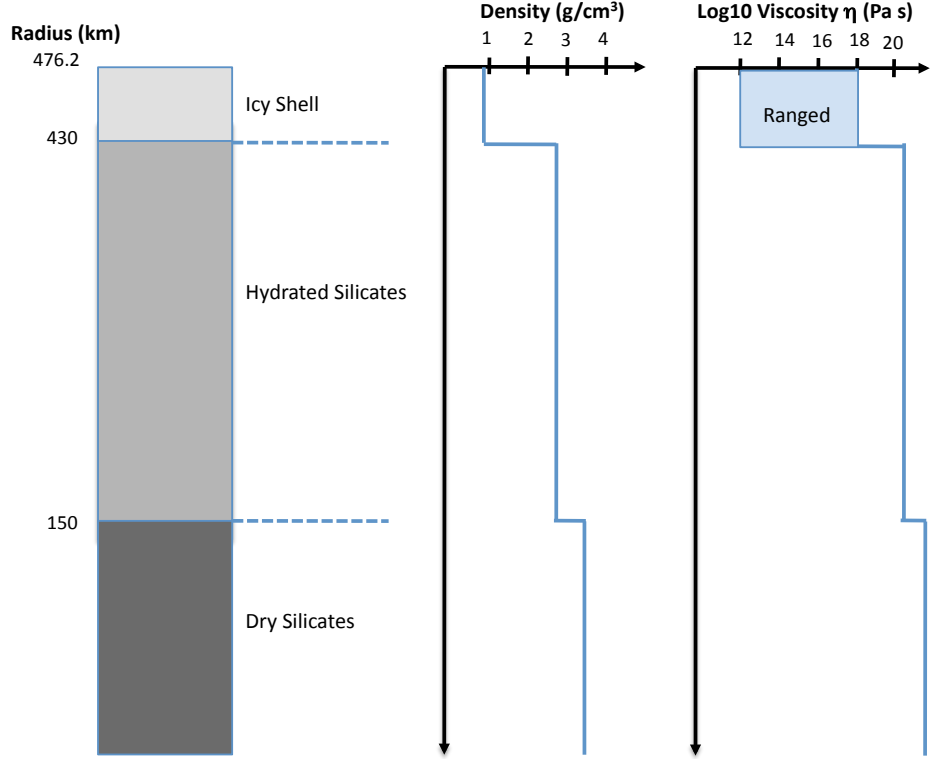}
\caption{Interior model of Ceres used in this study. The panels show, from left to right, the petrological structure, the corresponding density profile, and the viscosities assumed in the different layers.}
\label{Ceres}
\end{center}
\end{figure*}

Several shape models have been suggested for the past decade, inferred from different measurement techniques. Although the data are globally consistent, they show some discrepancy, as summarized in Zolotov (2009) and Castillo-Rogez and McCord (2010). The difference is in part due to the surface coverage enabled by the various techniques. As pointed out by Rivkin and Volquardsen (2009), longitudinal variations in the surface composition are likely to induce a bias in the interpretation of optical images obtained over a short longitudinal range. The difference between the $(a-c)$ radii difference (where $(a)$ and $(c)$ are the equatorial and polar radii, respectively)  inferred by Thomas \etal (2005) and Carry \etal (2008) is significant, of the order of 8 km, i.e., beyond the error bars of ~2 km estimated in both cases. The difference between the equatorial $(a)$ and the polar ($c)$ radii varies from 31.5 to 35.5 km. The upper bound indicates that CeresÕ  interior does not present large contrasts in density with depth. Zolotov (2009) inferred from that observation that Ceres is not chemically differentiated and that a small density gradient is due to the variation of porosity with depth, while Castillo-Rogez and McCord (2010) showed that even a warm icy satellite model whose core is dominated by hydrated silicate is consistent with that upper bound. A smaller value of $(a - c)$ is the signature of an increasing concentration with depth, for example due to the presence of an inner core composed of dry silicates (ordinary chondrite-like composition) that did not evolve since accretion or result from the dehydration of hydrated silicates (Castillo-Rogez and McCord 2010). 
In order to compute the rigid and non-rigid response of Ceres to external perturbations, we assume that the asteroid is stratified in a rocky and icy shell, after Castillo-Rogez and McCord (2010). Zolotov (2009) suggested that Ceres is a porous assemblage of hydrated minerals. However, Castillo-Rogez (submitted) demonstrated that hydrated minerals dehydrate in response to the moderate temperature increase undergone by Ceres in the course of its evolution. The main characteristics of this interior model are summarized in Fig.~\ref{Ceres}, and the geophysical parameters tested in this study are presented in Table~\ref{tab:param}.

\begin{table*}[htdp]
\begin{center}
\begin{tabular}{rrrrr}
Layer & Thickness & Density& Viscosity  & Shear Modulus\cr
&(km) & (kg/m$^{3}$) & $\eta$ (Pa s)& $\mu$ (GPa)  \\
\hline
Dry Silicate Core&$0-275$ & 3300 & 10$^{21}$ &  30   \cr
Hydrated Silicate Core&$125-414$ & 2700 &  10$^{20}$  &     30   \cr
Outer Icy Shell &$25-70$ & 931 & 10$^{12}$-10$^{19}$ &  3.3 
\end{tabular}
\end{center}
\caption{Main parameters ranged in the models tested for this study for the stratified model (Fig.~\ref{Ceres}). Ceres' mean radius and density are taken as 476.2 km and 2078 kg/m$^{3}$, respectively.}
\label{tab:param}
\end{table*}%

The mean moment of inertia is computed from the density profile after:
\begin{equation}
I  =   \frac{8\pi}{3} \int_{V_{body}} \rho(r) r^4 dr 
\end{equation}
where $V_{body}$ is the volume of the body, $\rho(r)$ is the density inside the body as function to the radius of the point $r$.  
The geophysical information contained in the global shape and degree-two gravity coefficients is a function of the secular tidal Love numbers $k_s$ (Munk and MacDonald 1960, and defined in Eq.~(\ref{eq:ks}))
\begin{equation}
J_2  = \frac{1}{3} m k_s \left(\frac{R}{a} \right)^{2} 
\label{eq:J2_I}
\end{equation}
where $m$ is the rotational parameter equal  to $\Omega^2 R^3 / \mathcal{G}M$, with $\Omega$ is the angular rotation rate, $\mathcal{G}$ the gravitational constant, $M$ the asteroid mass, and $R$ the mean radius of Ceres. 
In order to relate the internal structure to the observables, we can used the Radau-Darwin relationship written, in its approximated form, as (Van Hoolst \etal 2008): 
\begin{equation}
\frac{I}{MR^{2}} =  \frac{2}{3}\left(1-\frac{2}{5}\sqrt{\frac{4-k_s}{4+k_s}}\right)  
\end{equation}
\tnr{where the body is assumed to be in hydrostatic equilibrium}. We can then calculate the values of the equatorial and polar moments of inertia $A$ and $C$ from:
\begin{eqnarray}
\frac{C}{MR^2} & = & \frac{I}{MR^2} + \frac{2}{3}J_2\\
\frac{A}{MR^2} & = & - J_2 + \frac{C}{MR^2} 
\end{eqnarray}
where $J_2$ is the degree-two gravity coefficients. 
For the model presented in Figure \ref{Ceres} we obtain a mean moment of inertia $I/MR^2$ equal to $0.347$ and $A/MR^2=B/MR^2=0.3394$, $C/MR^2=0.3623$.
 
We compute the complex tidal Love number $k_{2}$, from the integration of the equations of motion by, e.g., Takeuchi and Saito (1972) (see Tobie \etal (2005) and Castillo-Rogez \etal (2011) for details about the computational approach.) Ceres' dissipation factor is inferred from the imaginary part of $k_2$. Mechanical attenuation is computed after the composite dissipation law introduced by Castillo-Rogez et al. (2011). That model is based on the observation that the attenuation spectrum of planetary materials shows a major shift in frequency-dependence as a function of the Maxwell time $\tau_M$ characterizing these materials (ratio of the viscosity to the shear modulus). At forcing frequencies greater than $2\pi/ \tau_M$, the dependence of the dissipation factor $Q$ on the angular frequency $\chi$ is such that $Q^{-1} \sim \chi^{-\gamma}$ with $\gamma = 0.2 - 0.4$. At low frequencies, the dissipation factor follows a Maxwellian behavior such that  $\gamma= 1$. This change reflects an evolution in the microstructural mechanisms driving dissipation: anelasticity-driven at high frequency and viscosity-driven at low frequency.\footnote{Anelastic strain is recoverable, but is a source of internal friction as it involves the motion of lattice defects. Viscoelastic involves the same defects, but is not recoverable.The anelasticity of planetary materials has been much studied, and a review can be found in McCarthy and Castillo-Rogez (2011).}  Castillo-Rogez et al. (2011) parameterized anelasticity using the Andrade model, with application to Iapetus, an icy satellite subject to a tidal stressing of a few kPa. The tidal stress amplitude in Ceres is of the order of 100 Pa, thus {\it{a priori}} we expect that the response to that stress involves the same physical mechanisms as described by Castillo-Rogez \etal(2011) for Iapetus. 
For the parameters displayed  in Table \ref{tab:param}, we find $k_2$ of the order of $10^{-3}$. Considering the absence of robust constraints on Ceres' temperature profile, a detailed calculation of the dissipation factor is meaningless. However, that parameter can be roughly quantified as a function of frequency. Indeed, Castillo-Rogez \etal (2011) demonstrated that \tjc{the dissipation factor of a water-rich object tends toward 1 at Ceres' orbital period (1681 days), but can significantly become greater than 100 at forcing periods as short as Ceres' spin period.} 

\section{Ceres' Pole Position}\label{sec:obliq}

\begin{figure}[htbp]
\begin{center}
\includegraphics[width=10cm]{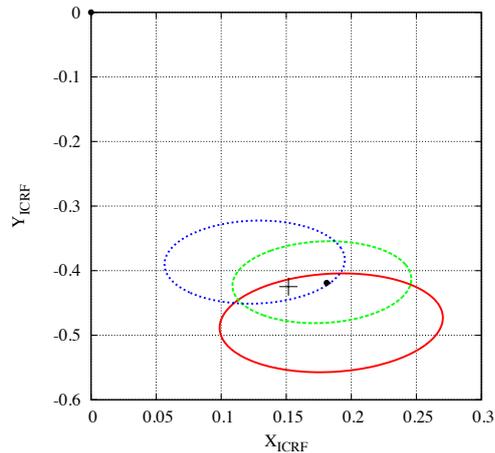}
\caption{Projection of the positions of the pole of rotation in the ICRF plane inferred by Thomas \etal (2005), red line; Carry \etal (2008), blue dotted line; Drummond and Christou (2008), green shaded line, from space telescope and ground-based measurements. The black dot represents the position of the orbital pole (Giorgini \etal 1996). The cross represents the mean intersection of the three measurements and the ellipses the uncertainties.}
\label{fig:obs_pole}
\end{center}
\end{figure}

\subsection{Polar orientation}\label{sec:pole_obs}
The precession-nutation theory of Ceres is defined for a given pole position of Ceres in space. However, the determination of the pole position of Ceres is difficult because of the size of the object and the lack of outstanding spectral features at its surface. Therefore, in this first section, we use the available observations based on adaptive optics and {\it{HST}} (Thomas \etal 2005; Drummond and Christou 2008; Carry \etal 2008) to address Ceres' pole position. The orientation data for Ceres' pole are provided in the International Celestial Reference Frame (ICRF) and are listed in the Table~\ref{tab:pole}. The first column represents the right ascension $\alpha_s$, the second column the declination $\delta_s$, and the last column the error bars. The Figure~\ref{fig:obs_pole} displays the three spin pole determinations of Table~\ref{tab:pole} projected onto the $XY$ plane of the ICRF, i.e. $X_{ICRF}$ and $Y_{ICRF}$:
\begin{eqnarray}
X_{ICRF} & = &\cos{\delta} \cos{\alpha}, \\
Y_{ICRF} & = &\cos{\delta} \sin{\alpha}.
\end{eqnarray} 
The three determinations overlap and their intersection is centered around $\alpha_s$=289.658 deg, $\delta_s$=63.189 deg, i.e. $(X_{ICRF}=0.1517,Y_{ICRF}=-0.4248)$, indicated by a cross in the Figure~\ref{fig:obs_pole}. 

\begin{table}[htdp]
\caption{Polar orientation of Ceres in right ascension ($\alpha_s$), declination ($\delta_s$) and uncertainties ($\Delta$).}
\begin{center}
\begin{tabular}{lrrr}
References & $\alpha_s$ (deg) & $\delta_s$ (deg) & $\Delta$ (deg) \\
\hline
Thomas \etal 2005 & 291  & 59  & 5 \\
Drummond and Christou 2008 & 293  & 63 & 4\\
Carry \etal 2008 & 288  & 66  & 5 \\
& \\
\multicolumn{4}{c}{Orbital orientation} \\
	 & $\alpha_n$ (deg) & $\delta_n$ (deg) &  \\
\hline
Horizons (Giorgini \etal 1996) & 293.39  & 62.85  & \\
\end{tabular}
\end{center}
\label{tab:pole}
\end{table}%

\subsection{Obliquity}

The Figure \ref{fig:obs_pole} shows the position of the orbital pole (black point) computed from the \textit{Horizons} \tnr{ephemerides} (Giorgini \etal 1996) and listed at the last line of the Table~\ref{tab:pole}. The orbital pole location coincides with the mean value of the pole position from Drummond \etal (2008). It is then interesting to investigate the information contained in the obliquity value, as previously suggested by Bills and Nimmo (2010). 
The obliquity $\eps$ is defined as the angle between the normal to the orbital plane and the figure axis of Ceres. 
If Ceres' obliquity has reached its equilibrium position as a consequence of internal dissipation, it is possible to obtain a relationship between the obliquity and the moment of inertia known as a Cassini state (e.g. Yoder 1995):
\begin{equation}
\frac{\nu}{n} \sin{(\eps-I)} = - \frac{3}{4}\sin{2 \eps} \frac{C-A}{C}
\label{eq:cassini}
\end{equation}
where $\nu$ = -50.48 kyr and $I $= 10.6 deg are the precession period and inclination of the orbit of Ceres with respect to the ecliptic plane (which is taken coincident with the Laplace plane), $n$ is the mean motion, and $A,C$ are the moments of inertia of Ceres (here, $A=B<C$). This formulae comes from the generalized Cassini states that result from an equilibrium position of the spin axis by taking into account the precessing orbit of the body (Colombo 1966; Peale 1969; Henrard and Murigande 1987; Lemaitre \etal 2006). By using a simple uniform precessional orbital period at 50.48 ky (Bills \& Nimmo 2010) and our model of Ceres differentiated into a rocky core and icy shell (Section~\ref{sec:shape}), we obtain an equilibrium obliquity of $\sim$0.01 deg. This small value is mainly due to the long precession period ($\sim$-50.48 kyr) with respect to the orbital period (1681 days). The secular motion of Ceres is influenced by the oscillation at -22 kyr (see Bills \& Nimmo 2010) and in this case the equilibrium obliquity is equal to 0.02 deg. Relaxation to the Cassini state is achieved when the obliquity $\eps$ meets this equilibrium criterion. 

\begin{figure}[htbp]
\begin{center}
\includegraphics[width=15cm]{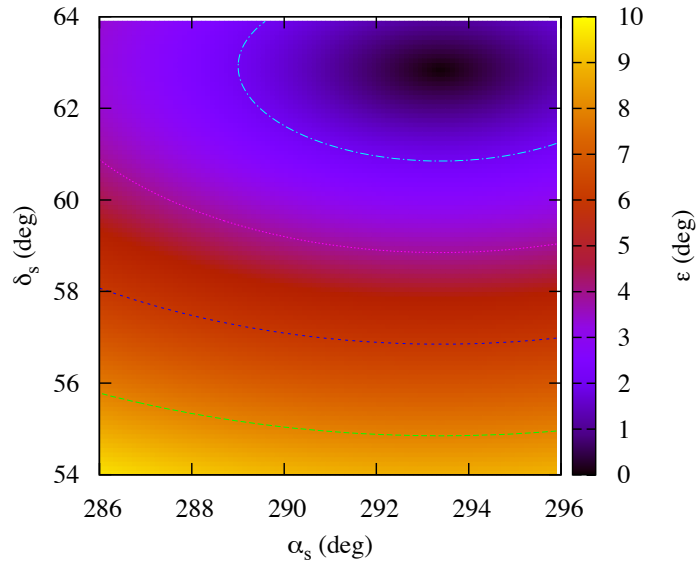}
\caption{Obliquity of Ceres as a function of the orientation of its spin orientation in the equatorial reference frame. The nominal value detected by Thomas \etal (2005) is  $\alpha_s = 291 \pm 5$ deg and $\beta_s=59 \pm 5$ deg. \tnr{The curves represent isocontours every 2 degrees.}}
\label{fig:obliquity}
\end{center}
\end{figure}

From the right ascension $\alpha_n$ and declination $\delta_n$ of the orbit pole, we could express the obliquity as 
\begin{equation}
\cos{\eps} = \sin{\delta_s}\sin{\delta_n} + \cos{\delta_s}\cos{\delta_n}\cos{(\alpha_s - \alpha_n)}
\end{equation}
To compute the orbit pole coordinates, we use the $Horizons$ ephemeries that provides the orbital coordinates in the ecliptic reference frame (the orbital inclination of Ceres and the ascending node are 10.6 deg and 80.5 deg, respectively). Then we express these coordinates in the ICRF by using the Earth's obliquity. The final coordinates of the orbital pole are reported in Table~\ref{tab:pole}. 
The obliquity is equal to 4.01 degrees for the mean pole orientation of Thomas \etal (2005), 0.23 deg for Drummond and Christou (2008), and 3.91 deg for Carry \etal (2008). The observation of Drummond and Christou (2008) seems to be close to that expected if Ceres is relaxed to a Cassini state. The uncertainty on the obliquity is represented in Figure~\ref{fig:obliquity} in the case of the measurement obtained by  Thomas \etal (2005). The obliquity is between 0 and 10 degrees i.e. that contains the Cassini state but present very large variation.  
For this paper, we use as a working reference for Ceres' orientation the upper value of the obliquity of 9.6 deg. The reason for using that large value, while data overlap for an obliquity value of $\sim$ 3 deg, is that it will yield upper bounds on our estimates of the rotational perturbations. This will help us to assess whether or not these perturbations can be measured with spaceborne techniques.

Bills and Nimmo (2010) predicted that the obliquity of Ceres is $\sim$12 deg based on a secular orbital model of Ceres. However, available observations and the location of the orbital pole yielded by the $Horizons$ ephemeris indicate that the obliquity is most likely between 0 and 10 deg and consistent with the conclusion of Thomas \etal (2005) that the obliquity of Ceres is around $3 \sim 4$ deg. The discrepancy seems to reside in the initial value used in Eq.~(20) of Bills and Nimmo (2010) and it could be resolved by using the initial values output by the $Horizons$ ephemerides.
 
\subsection{Damped obliquity}

We now evaluate to what extent Ceres has evolved toward that equilibrium state. 
The obliquity damping rate $\dot{\eps}$ may be computed from the following equations (e.g., N\'eron de Surgy \etal 1996; Levrard \etal 2007) describing the secular rotational evolution of Ceres (for an orbit without planetary perturbations):
\begin{eqnarray}
\dot{\eps}  &= &\frac{K n }{C \Omega} \sin{\eps}  [ \cos{\eps} \; f_1(e) \frac{\Omega}{2n} - f_2(e)]  \\ 
\dot{\Omega} &= &- \frac{K n }{C} [ f_1(e) \frac{1+\cos{\eps}^2}{2} \frac{\Omega}{n} -  f_2(e) \cos{\epsilon}]
\label{eq:dissipeta}
\end{eqnarray}
where
\begin{eqnarray}
f_1(e)  =  \frac{1+3e^2+3e^4/8}{(1-e^2)^{9/2}} \;\; , \;\; f_2(e)  =  \frac{1+15e^2/2+45e^4/8+5e^6/16}{(1-e^2)^{6}}
\end{eqnarray}
where $e$ is the eccentricity and the constant $K$ is defined as
\begin{equation}
K=3\frac{k_2}{Q}\frac{ \mathcal{G}M^{2}}{R_e} \left (\frac{M_{\odot}}{M}\right )^2 \left ( \frac{R_e}{D}\right )^6 n 
\end{equation}
with $\mathcal{G}$ is the gravitational constant, $M_{\odot}$ the mass of the Sun, $D$ Ceres' semi-major axis, $R_e$ its equatorial radius, and $\Omega$ the angular rotation rate. The parameters $k_2$ and $Q$ correspond to the tidal Love number and dissipation factor, respectively \tnr{at the orbital frequency. For the sake of simplicity, we assume for this calculation that these parameters remain constant}. The inverse dependence of that equation on $D^6$ suggests that Ceres' large distance to the Sun is a severe limitation to any tidal evolution of its dynamical properties. In addition, in this expression, we neglect the impact of orbital perturbations. 

From Equations~(\ref{eq:dissipeta}), as quoted by Correia (2009) the timescale of evolution of the rotation rate is shorter than the timescale for the obliquity evolution. So it is expected that first the rotation reaches its equilibrium that will take a damping rate of the order of  
\begin{equation}
\dot{\Omega} \sim  6.14 ^{-11}\frac{k_2}{Q} \;\; , \;\; \dot{\epsilon} \sim  1.00 ^{-14}\frac{k_2}{Q}
\end{equation}
where time is expressed in years, and then the obliquity will reach its equilibrium state.  
The parameters $k_2$ and $Q$ are not known but may be approached from geophysical modeling. 
Let us note that Bills and Nimmo (2010) considered a situation where Ceres' material is in equilibrium at the dissipation peak. However, such a situation is less than likely. The contribution of tidal dissipation to the total heat budget of the object is negligible (below $0.1\%$) in comparison to insolation and, to a lesser extent, long-lived radioisotopes. This precludes that heat source to drive the geophysical state of the asteroid. For the reference model chosen for this study (detailed in the previous Section), differentiated in an icy shell and rocky core,  $k_2$ is of the order of $10^{-3}$ and $Q$ is of the order of 10. 
\tnr{In that case the dissipation time is of the order of 10$^{13}$ years for $\dot{\Omega}$ and 10$^{17}$ years for $\dot{\epsilon}$.} Hence we infer that Ceres' obliquity has probably not fully damped over its lifetime. As a consequence, our result contrasts with the conclusion of Bills and Nimmo (2010) that Ceres' obliquity could have damped in a few hundred My, even if we assume that the asteroid is in a very dissipative state. 


\section{Rotational model of Ceres}\label{sec:rotation}

\subsection{Euler-Liouville Equations}

If Ceres were perfectly spherical and rigid, then its rotation would be uniform. However, the Hubble Space Telescope measurements showed that the figure of Ceres is an oblate body (within the error bars), from which we inferred an equatorial oblateness $\alpha = (C-A)/A = 0.0675$ 
(see section~\ref{sec:shape}). Thus, the Sun exerts a non-zero torque on Ceres dynamical figure, which responds through precession and nutation of its orientation axes. The Sun also raises tides that deform its surface and perturb its rotational velocity. 

Thus it is convenient to describe Ceres' rotation by using the approach developed for Earth, which is oblate at first order. The rotation of the body is described through the classical Euler-Liouville equation written as (see Moritz \& Mueller 1987;  Dehant and Mathews 2007) 
\begin{equation}
\frac{d \vec H}{dt} + \vec \Omega \wedge \vec H = \vec \Gamma
\label{eq:eurlerliouville}
\end{equation}
This describes the variations of the angular momentum $\vec H$ disturbed by an external torque $\vec \Gamma$. This equation is expressed in the rotating frame tied to the body through the spin velocity $\vec \Omega$ and written in the Tisserand frame (Munk and MacDonald 1960). As shown in the section (Section~\ref{sec:geocons}) the wobble damping time is some ten thousands of years 
and we could assume that the instantaneous axis of rotation is near the polar principal axis of the body. Thus,
\begin{equation}
 \vec \Omega = \left( 
\begin{array}{c}
m_1 \\
m_2\\
1+m_3\\
\end{array}
\right) \Omega
\end{equation}
where $\Omega$ is the mean rotation of the body and the quantities $m_i$ are small and dimensionless. The pair $m_1, m_2$ describes the polar motion of Ceres, i.e. the orientation of the rotational speed in the body reference frame, while $m_3$ corresponds the variation in the rotational speed as shown in the following linearized expression
\begin{equation}
 || \vec \Omega || = \Omega(1+m_3).
\end{equation}

The angular momentum is $\vec H = I \vec \Omega$ with $I$ the tensor of inertia of the body expressed as 
\begin{equation}
 I = \left( 
\begin{array}{ccc}
A & 0 & 0 \\
0 & A & 0 \\
0 & 0 & C \\
\end{array} 
\right) +  
\left( 
\begin{array}{ccc}
c_{11} & c_{12} & c_{13}\\
c_{21} & c_{22} & c_{23} \\
c_{31} & c_{32} & c_{33} \\
\end{array} 
\right) 
\label{eq:inertia}
\end{equation}
where the $c_{ij}$ are symmetric and represents the departure from the reference ellipsoid, i.e. the deformation of the body surface. By introducing the moment of inertia Eq.~(\ref{eq:inertia}) into the dynamical equations Eq.~(\ref{eq:eurlerliouville}) and developing at first order in $m_j$ and $c_{ij}$, the linearized dynamical are
\begin{eqnarray}
A \Omega \dot{m_1} + (C-A) \Omega^2 m_2 + \Omega \dot{c}_{13} - \Omega^2 c_{23} = L_1 \nonumber \\
A \Omega \dot{m_2} - (C-A) \Omega^2 m_1 + \Omega \dot{c}_{23} + \Omega^2 c_{13} = L_2 \\
C \Omega \dot{m_3}  + \Omega \dot{c}_{33}  = L_3 \nonumber
\label{eq:firstorder}
\end{eqnarray}
or by introducing complex notations, as usual for Earth rotation studies, $m=m_1+i m_2$, $L=L_1+i L_2$, $c=c_{13}+i c_{23}$, we obtain one complex equation for the polar motion
\begin{equation}
A \dot{m} - i \alpha A \Omega m + \dot{c} + i \Omega c = \frac{L}{\Omega}
\label{eq:polarmotion}
\end{equation}
and one equation for the l.o.d (length of day) variations 
\begin{equation}
C \Omega \dot{m_3}  + \Omega \dot{c}_{33}  = L_3
\label{eq:m3L3}
\end{equation}
It is interesting to note that, at the order of approximation of axi-symmetric body, the polar motion and l.o.d variation are described by well separated dynamical equations. 

\subsection{Deformed tensor of inertia}
The centrifugal and tidal potentials deform the body and that deformation may be expressed as (Dehant \etal 2005)
\begin{equation}
c = \alpha A \frac{k_2}{k_s} m - 3\alpha A \frac{k_2}{k_s}\frac{W_{21}}{\Omega^2 d^2}
\end{equation}
where the first term results from the centrifugal potential and the second one from the tidal potential and especially the tesseral potential  $W_{21}$ (see Section~\ref{sec:potential}). $G$ is the gravitational constant, $d$ is the mean Sun-Ceres distance, $k_2$ is the tidal Love number, and $k_s$ the secular Love number 
defined as (Munk and MacDonald 1960): 
\begin{equation}
k_s = 3\frac{(C - A)\mathcal{G}}{\Omega^2 R^5}
\label{eq:ks}
\end{equation}
where $R$ the radius of the surface.

The $c_{33}$ tensor varies as a function of the centrifugal and tidal potentials after (Greff-Lefftz \etal 2000)
\begin{equation}
c_{33} = -\frac{4}{3} \alpha A \frac{k_2}{k_s} m_3 - 2\alpha C \frac{k_2}{k_s}\frac{W_{20}}{\Omega^2 d^2}
\label{eq:c33}
\end{equation}
where $W_{20}$ is the zonal potential (see Section~\ref{sec:potential}). 

\subsection{Gravitational torque}

The tesseral degree 2 tidal potential $W$ acts on the Ceres' equatorial bulge and thus involves an equatorial torque such that \tnr{(Dehant and Mathews 2007)}
\begin{equation}
L_1 +i L_2 = -\frac{3i\alpha A }{d^2} W_{21}
\end{equation}
where the complex potential $W_{21}$ is developed in the next section. The torque $L_3$ is equal to zero due to the symmetry axis of Ceres.

\subsection{The tidal potential}\label{sec:potential}

The gravitational tidal potential induced by the Sun may be expressed in both, a frame tied to Ceres (MBRF=Mean Body Reference Frame) and the celestial frame (MCRF=Mean Celestial Reference Frame). Choosing one reference frame, phenomena induced by the gravitational forcing have to be expressed in the same frame, with the particularities that the frequency in the frame tied to Ceres and the frequency in space are related by the rotational velocity of the body (see Eq.~(\ref{eq:chgtrepere})). So a constant torque applied in the MCRF will appear periodical in the MBRF at the rotational frequency and {\it{vice versa}}.   

\tnr{Following, the method of (Dehant and Mathews 2007), the degree-two potential exerted by the Sun on Ceres in the MBRF is developed as
$W_2 = W_{20} + W_{21} $
\begin{equation}
W_{20} = \frac{\mathcal{G}M}{d^3} (z^2 - \frac{1}{3})
\end{equation}
and 
\begin{equation}
W_{21} = \frac{\mathcal{G}M}{d^3} (xz + i yz)
\end{equation}
where the $W_{20}$ leads to the zonal part and $W_{21}$ leads to the sectorial part in the tidal torque. 
Here, the sectorial} part is zero due to the axi-symmetry shape of the body. The cosine directions $(x,y,z)$ are the direction of the Sun in the MBRF. They are evaluated from the Horizons ephemeris (Giorgini \etal 1996) and rotate from the ecliptic frame to the MBRF by using the polar direction of Ceres with a right ascension of 286 deg and declination of 54 deg consistent with an obliquity of 9.2 deg. The uncertainty related to the direction of the polar direction of Ceres is discussed in section~\ref{sec:pole_obs}. 

\tnr{The zonal part is developed as a Fourier series $W_{20} = W_{20}^j e^{i \omega_j t} $ where $W_{20}^j$ contains also the phases} and the tesseral part is developed as Fourier series according prograde (index~$+$) and retrograde (index~$-$) components: 
\begin{eqnarray}
W_{21} = W_{21}^+ e^{i \omega_j t} + W_{21}^- e^{i \omega_{-j} t}
\label{eq:potential}
\end{eqnarray}
where we used the same notations as (Roosbeek 1995; Roosbeek and Dehant 1998). The prograde and retrograde circular motions allow to express the elliptical motion in two symmetric motion components. The frequencies $\omega_j$ of the tidal potential are expressed in the MBRF, and they are related to the prograde and retrograde frequencies ($\Delta \omega_j, -\Delta \omega_j$) expressed in the space MCRF, through the following relation: 
\begin{eqnarray}
\omega_j = \Delta \omega_j - \Omega \nonumber \\
\omega_{-j} = -\Delta \omega_j - \Omega 
\label{eq:chgtrepere}
\end{eqnarray}
where $\Omega$ is the rotation period close to 9 hours. Henceforth, the periods appear long-period in space and short period in the frame tied to Ceres.

In the case of the Earth, one goes from the terrestrial reference frame tied to the planet to the celestial frame using several rotation matrices. These rotations first bring the terrestrial frame attached to the figure axis of the Earth to the intermediate pole, accounting for polar motion; then a rotation is performed along the true equator of date around this intermediate pole, accounting for the Earth's rotation (uniform part and length-of-day or UT1 variations). Then the precession and nutations are accounted for in order to bring the true equator of date to the celestial frame. Precession and nutations are thus those of the true equator of date. One has to keep in mind however that the choice concerning the intermediate frame is purely conventional. The more logical choice is of course related to the way UT1 or the length-of-day variations are expressed. In the recently adopted conventions for Earth, the intermediate frame is the equator of the CIP (Celestial Intermediate Pole), a conventional pole that has no retrograde diurnal motion in a reference frame tied to the Earth, and only long-period motions (precession, nutations) in space.
The instantaneous rotation pole and the mean rotation pole are not identical; they differ by small changes in their direction due to atmosphere, ocean and hydrology excitation of polar motion at very short periods. When computing the precession and nutations, these axes have identical long-period motion in space and retrograde diurnal motion in a frame tied to the Earth. 
Similarly, for Ceres, we can ignore the differences at short period in space. We then work with the instantaneous rotation pole $(m_1, m_2)$ in a frame tied to Ceres. Long-term motion in that frame will be related to the Chandler Wobble, if excited. Long-term motion of the pole in space or retrograde diurnal motions in the frame tied to Ceres are representing precession and nutations. We have a one-to-one relation between the frequencies of these motions in a frame tied to Ceres and in space.

\section{Description of the rigid rotational motion}\label{sec:rigide}

\subsection{Polar motion}

\begin{table*}[htdp]
\caption{Ceres rigid Polar motion ($k_2=0$). The prograde motions are shown on the left columns of the Table whereas the right part of the Table contains the retrograde motions. 
}
\begin{center}
\begin{tabular}{rrr|rrr}
 Freq $\omega_j$ & Per  & Amp    & Freq $\omega_{-j}$ & Per  & Amp     \cr
 (rad/days) &  (days) & (mas)   &  (rad/days) &  (days) & (mas)    \cr
\hline
\hline
     -16.617 &     -0.37813 &      0.1683 &    \cr 
     -16.613 &     -0.37821 &      0.0299 &    -16.620 &     -0.3780 &     -0.0101  \cr 
     -16.609 &     -0.37830 &      0.1638 &    -16.624 &     -0.3780 &      0.0011  \cr 
     -16.605 &     -0.37838 &      0.0449 &    -16.628 &     -0.3779 &     -0.0003  \cr 
     -16.605 &     -0.37840 &      0.0009 &    -16.629 &     -0.3779 &      0.0000  \cr 
     -16.602 &     -0.37847 &      0.0085 &    -16.632 &     -0.3778 &     -0.0001  \cr 
     -16.598 &     -0.37855 &      0.0014 &    -16.635 &     -0.3777 &      0.0000  \cr 
\end{tabular}
\end{center}
\label{tab:polarmotion}
\end{table*}%

First, we solve the polar motion by introducing the potential expressed in~(\ref{eq:potential}) and using a Fourier transform $e^{i\sigma t}$ to express the budget equations at a given frequency in Equation~(\ref{eq:polarmotion}). The complex polar motion is then: 
\begin{eqnarray}
m & = & m_c e^{i \sigma_c t } e^{-\lambda t}  + \\
& & \sum_j \big[
\frac{-3\alpha W_{21}^+}{\Omega^2 d^2 } 
\frac{\Omega - (\Omega + \omega_j) k_2/k_s }
	{\omega_j - \Omega \alpha + \alpha (\Omega+\omega_j)k_2/k_s }
	e^{i (\omega_j t + \phi_j)} + \nonumber \\
& & \frac{-3\alpha W_{21}^-}{\Omega^2 d^2 } 
\frac{\Omega - (\Omega + \omega_{-j}) k_2/k_s }
	{\omega_{-j} - \Omega \alpha + \alpha (\Omega+\omega_{-j})k_2/k_s }
	e^{i (\omega_{-j} t + \phi_j)} \big] \nonumber 
\end{eqnarray}
hence the polar motion is composed of a free mode (first term) and a sum of forced modes, prograde and retrograde. 

The frequency of the free mode is called the Chandler frequency $\sigma_c$ by analogy with Earth rotation and is written as 
\begin{equation}
\sigma_c =\alpha\Omega \frac{k_s^2-|k_2| k_s \cos{\delta} + |k_2| \alpha( k_s \cos{\delta} - |k_2|)}{k_s^2+2\alpha |k_2|k_s \cos{\delta}+\alpha^2 |k_2|^2}
\end{equation}
Its period is about $5.48$ days and the correction due to the deformation \tnr{$k_2 = |k_2| e^{(-i \delta)}$ with $\delta$ the phase lag representing the dissipative part,}  
is between 3 and 5\% of its value. 
This contrast to the Earth's case, for which that deformation induces a difference of 100 days in the period. The difference of behavior comes mainly from the value of $\alpha$ that differs by about one order of magnitude between the two bodies. The amplitude $m_c$ and the phase $\phi_c$ of the Chandler mode depend on the dynamical and geophysical history of the body (see discussion section~\ref{sec:excitwobble}). The amplitude of the Chandler mode is damped with a typical timescale $1/\lambda$, a function of the imaginary part of the Love number and expressed as 
\begin{equation}
\lambda = \frac{\alpha\Omega |k_2|k_s \sin{\delta}(\alpha+1)}{k_s^2+2\alpha |k_2|k_s\cos{\delta}+\alpha^2 |k_2|^2}
\label{eq:lambda}
\end{equation}
Its value strongly depends on the interior model as discussed in section~\ref{sec:geocons}.

\begin{figure}[htbp]
\begin{center}
\includegraphics[height=10cm]{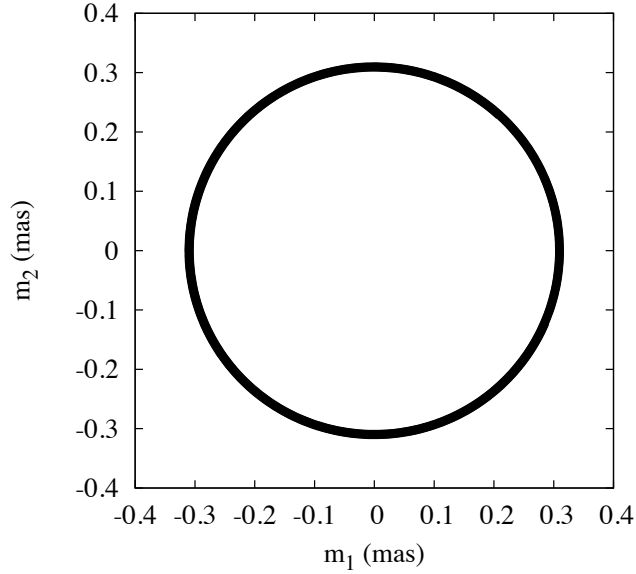}
\caption{The polar motion of Ceres is circular and present a main term oscillating at 840 days. 
}
\label{fig:polarmotion}
\end{center}
\end{figure}

The forced terms of the rigid Ceres' polar motion are shown in Table~\ref{tab:polarmotion}. The polar motion of Ceres oscillates at short periods close to 9 hours and its motion projected onto the surface of Ceres is very small (see Figure~\ref{fig:polarmotion}). Its amplitude multiplied by the mean radius of Ceres is $\sim$ 0.5 millimeter, and summing all the contributions regardless of the phase yields an amplitude no greater than 1 millimeter.

\subsection{Precession-Nutation of Ceres}

The rotational motion of Ceres' polar axis describes in the inertial reference frame, the MCRF, a precessional nutational motion. The rotation angles and their derivatives are easily computed by using the kinematic Euler equation allowing to express the instantaneous rotation pole components in terms of the nutation angles as
\begin{equation}
 \dot{\theta} + i \dot{\psi}\sin{\theta} = \Omega m e^{i \Omega t}
 \label{eq:kinematics_euler}
\end{equation}
where $\Omega$ accounts for the expression of the pole in space due to the rotation around the Z-axis. So after integration of Eq~(\ref{eq:kinematics_euler}), \tnr{except for the case $\omega_j=-\Omega$ that leads to the precessional motion, we obtain the following nutation series}
\begin{eqnarray}
\Delta \theta + i \Delta {\psi}\sin{\theta} & =  & -i \frac{\Omega}{\Omega + \sigma_c} m_c e^{i (\Omega +\sigma_c)t}e^{-\lambda t} \\
& & +i \sum_{j\neq0} \big[ \frac{- \Omega}{\Delta \omega_j} m_j^+ e^{i (\Delta \omega_j t + \phi_j)}  + \frac{ \Omega}{\Delta \omega_j} m_j^- e^{-i (\Delta \omega_j t + \phi_j)} \big] \nonumber
\label{eq:nutprec}
 \end{eqnarray}
The rigid nutations of Ceres are described by the periodic components of the last equation and they are listed in Table~\ref{tab:nutations}. The amplitudes of the long-period nutations are positively affected because the amplitude is inversely proportional to the forcing frequency. The main term (in absolute amplitude) is the \tnr{semi-annual nutation $2 \lambda_c$} related to the obliquity of Ceres, and then the terms related to harmonics. \tnr{We also note the presence of a term related} to Jupiter's mean longitude $\lambda_J$. The amplitude of the annual nutation is around 364 mas, that is a 0.84 m surface displacement, for a mean radius of 476 km. The detection of such a small displacement requires tracking of Ceres' surface with a beacon for long periods of time. 
\begin{table*}[htdp]
\caption{Ceres rigid Nutations ($k_2=0$) and corresponding argument with $\lambda_c$ and $\lambda_J$ mean longitudes of Ceres and jupiter. The amplitudes are truncated at $10^{-4}$ mas.}
\begin{center}
\begin{tabular}{lrrrr}
Arg &  Freq          & Per      & Amp Nut prog    & Amp Nut retro \cr
 &(rad/days) &  (days) &  (mas)   & (mas) \cr
\hline
\hline
$2 \lambda_c$   &       0.00747 &    840.8 &   -364.296  &      2.504 \cr 
$ \lambda_c$ &           0.00374 &   1681.7 &   -133.129  &    -44.976  \cr 
$3 \lambda_c$    &      0.01121 &    560.6 &    -66.627  &     -0.400  \cr 
$4 \lambda_c$      &    0.01495 &    420.4 &     -9.449  &     -0.054 \cr 
$4 \lambda_c-\lambda_J$ &       0.01204 &    521.7 &     -1.256  &      0.009 \cr 
$5 \lambda_c$   &       0.01868 &    336.3 &     -1.217 &        0.007 \cr 
\end{tabular}
\end{center}
\label{tab:nutations}
\end{table*}%

The first term in equation~(\ref{eq:nutprec}) represents the Chandler mode observed from space. In this case, it has a period of 9h40 minutes i.e. an increase of ~36 minutes with respect to the  proper rotation of the body. \tnr{The precessional motion of the figure axis is represented by} the oscillation at $\omega_j = -\Omega$ in the body reference frame that \tnr{is purely imaginary $m_0 =  0.1576 $ mas}, leading to a precession time of 226981.8 years that is \tnr{longer} than the 218654.2 years period calculated with the classical formulae: 
\begin{equation}
\dot{\psi} = -\frac{3}{2}\frac{n^2}{\Omega} \frac{C-A}{C }\cos{\epsilon}
\end{equation} 
The discrepancy between the two results (4$\%$) is essentially due to the ephemeris timescale used in the frequency analysis of the potential.

\section{Geophysical constraints from space observations}\label{sec:geocons}

\subsection{Non-rigid contributions: l.o.d}

In section~\ref{sec:rotation} we have introduced the rotational equations for a non-rigid body. The application of these equations to the geophysical models shows that the non-rigid contributions to the shape deformation bear a negligible effect on polar motion and nutational motion. However, the variations of the moments of inertia in response to the tidal forcing exerted on the body generates a non-zero torque along the figure axis that would perturb the uniform rotational motion in the form of length of day (l.o.d) variations. By combining Eq.~(\ref{eq:m3L3}) and the inertia deformation~(\ref{eq:c33}) we deduce the variations of the l.o.d $m_3$
\begin{equation}
m_3 = \frac{2 \alpha \frac{k_2}{k_s}}{1 - \frac{4}{3} \alpha  \frac{A}{C} \frac{k_2}{k_s}}\frac{W^j_{20}}{\Omega^2 d^2}
\end{equation}
The resulting oscillations of the $m_3$ variations are very small under 0.001 mas (Table~\ref{tab:m3}), largely below the expected accuracy for space-borne observational techniques. \tnr{The term at 1374.3 days is related to Jupiter with the combination $2\lambda_c - \lambda_J$.} 
\begin{table}[htdp]
\caption{$m_3$ variations of Ceres for a tidal love number $k_2=0.017-i \,1.27\;10^{-7}$ \tnr{with frequencies expressed in the inertial reference frame}. The amplitudes are truncated at $10^{-5}$ mas.}
\begin{center}
\begin{tabular}{rrrr}
 Freq          & Per      & Amp in-phase     & Amp out-of-phase \cr
 (rad/days) &  (days) &  (mas)   &  (mas)  \cr
\hline
\hline
       0.00374 &   1681.9 &      0.00072 &     0.0   \cr 
       0.00747 &    840.8 &       0.00020 &     0.0   \cr 
       0.01121 &    560.6 &       0.00004 &     0.0   \cr 
       0.00457 &   1374.3 &     0.00001&     0.0   \cr 
\end{tabular}
\end{center}
\label{tab:m3}
\end{table}%

\subsection{Wobble}\label{sec:excitwobble}

\begin{figure}[htbp]
\begin{center}
\hspace{-2cm}
\includegraphics[width=10cm]{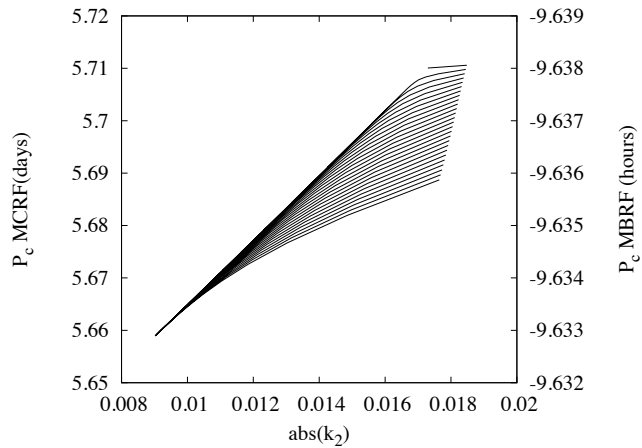}
\caption{Period of the Chandler model as a function of the value of the Love number $k_2$. The lower bound of $k_2$ corresponds to a fully frozen model while the upper bound is expected if Ceres contains a global ocean. }
\label{fig:chandler}
\end{center}
\end{figure}

The rotational motion of Ceres appears to be relatively uniform because all the nutational oscillations, polar motion, and l.o.d variations show very small amplitudes. Therefore, if a sizeable departure from a quiet rotation is detected by the $Dawn$ Mission at a period of about of ~9h40, then we could assign this motion to the Wobble. Indeed, the presence of a Chandler mode is expected as soon as any perturbation, exterior or interior to the body, shifts the figure axis from its equilibrium position. However, this mode is also damped due to internal dissipation.
The Chandler mode expressed in the inertial reference frame has a period of 9h40 minutes, i. e., around 36 minutes longer than the  proper rotation of the body. As the nutations in the inertial frame are of long periods (harmonics of the orbital period), then any observed departure of the uniform rotation at the short period of 9h40 might be attributed to the Chandler mode. The Chandler period is sensitive to the value of the Love number $k_2$ as shown in Figure~\ref{fig:chandler}, where the period is expressed in both reference frame MBRF (5.5 days) and MCRF (9h40). We consider a wide range of possible values for $k_2$ covering the spectrum of possible models envisioned for Ceres. This parameter is computed at the period of the Chandler mode at 5.5 days, i.e. the period in the body reference frame. 
The damping timescale $ T_\lambda$ is proportional to $Q$ and the damping timescale can be as long as 120,000 years for dissipative models with large $Q$ and as short as as few decades if the object is very dissipative. Thus the damping timescale {could be} very short, hence a non-zero Chandler mode requires a continuous physical process or a recent impulse in order to be observable today. 

In the Earth's case, the Chandler wobble is mainly excited by the atmosphere and the ocean. In the case of Ceres there is no atmosphere, but Ceres is in a rich dynamical environment, the asteroid belt, thus it is exposed to a constant meteoritic flux. 
Such meteoritic flux may involve impacts exciting the Chandler mode for Ceres. The wobble excitation may be expressed by using Peale (1975) expression introduced in the case of the Moon: 
\begin{equation}
m_e = - \frac{i c}{A \alpha} H(t-t_0)  + (-\frac{c}{A} + \frac{ic}{A\alpha} + \frac{N}{A\alpha} )H(t-t_0)e^{i\alpha\Omega(t-t0)} 
\label{eq:impact}
\end{equation}
where $m_e$ represents the polar response of the impact, $N$ represents the maximum angular momentum potentially induced on Ceres by a collision ($N=mvR$ with $m$ and $v$ corresponding to the mass and velocity of the bolide, respectively, and $R$ Ceres' radius ). The parameter $H(t-t_0)$ is the Heaviside function associated with an impact at time $t_0$. This expression contains two components: the angular momentum transfer and the modification of the moment of inertia due to the ejected matter and formation of a crater. The subsequent response of the pole is composed of a constant offset due to the first term in Eq.~(\ref{eq:impact}) and excitation of the Chandler mode. We use the formulation of Gauchez \& Souchay (2006) for crater modeling and the scaling law is borrowed from Holsapple (1993). We search for possible impact configurations leading to the excitation of the Chandler mode with an amplitude of 10 arcseconds, i.e., a displacement at the surface of 20 meters (the amplitude observed today would be damped due to the dissipation of the Chandler mode, so such events have to be recent). This may be achieved for a cometary projectile (heliocentric) with a diameter of 2.5 km, a density of 0.6 g/cm$^3$, and a velocity of 20 km/s; or by a neighbor asteroid of 4 km diameter with a density of 1.3 g/cm$^3$, colliding at 5 km/s (Farinella and Davis 1992). 

In order to estimate the probability of such an impact on Ceres, we survey main-belt asteroids with absolute magnitude lower than 14. The population contains approximately 25000 objects (see Jedicke \etal 2002). Trajectories of all the considered asteroids were calculated for a 100-year time interval assuming Keplerian orbits. A
fictional object evolving on the same orbit as Ceres, but with a cross-section 100,00 times greater would experience 200 collisions with other asteroids. Scaling this value to Ceres' size and a time
span of 150,000 years, we obtain 0.003 impacts on Ceres during that timeframe. An asteroid
diameter of 4 km corresponds approximately to an absolute magnitude of 15.
According to Jedicke \etal (2002), these objects are 2 to 3 times more
abundant than the population considered here. As a consequence, the corresponding number of
impacts onto Ceres amounts to approximately 0.007 per 150,000 yrs. The probability that Ceres experienced in the last 150,000 years a collision with an object
greater than 4 km appears to be small (less than 1$\%$). Such an estimate is more difficult to calculate in the case of cometary collisions due to the lack of constraints on the possible reservoir of comets. 

Another consequence of collision with large objects is the alteration of the moments of inertia of Ceres that may lead to shift of its figure axis (first term in Eq.~\ref{eq:impact}). The long-term consequence of that effect needs to be studied in details. It requires to properly model the respective timescales for the relaxation of the crater and of the equatorial bulge. As noted by Nimmo and Matsuyama (2007), both processes rely on the mechanical properties of the icy shell and thus should proceed over the same timeframe, which increases the complexity of the problem. For Ceres, the low subsurface viscosity should promote rapid crater relaxation preventing the re-orientation. 

We also checked for the possible occurrence of close encounters during the \textit{Dawn} mission lifetime, which could excite Ceres' spin axis by an impulse of its gravitational torque. From realistic (non keplerian) asteroid orbits, we found no close encounter between 2010 and 2020 that could modify the rotational dynamics of Ceres. The encounters are not sufficiently close or the bodies involved are not massive enough. This estimate accounts only for main-belt asteroids with absolute magnitudes lower than 14.

Another source of excitation of the Chandler wobble may be due to the presence of an equatorial sea inside Ceres. The existence of such a water reservoir has been suggested by Castillo-Rogez and McCord (2010) based on the observation that Ceres' surface temperature at the equator is close to the eutectic temperature of salt impurities expected in the asteroid. Several recent astrophysical models also suggest that Ceres accreted a significant fraction of ammonia hydrates (up to  7wt.\% of the ice phase, Dodson-Robinson \etal 2009), and possibly also methanol hydrates (Mousis \etal 2008). The presence of these compounds would help preserve a deep liquid layer over extended periods of time, as, e.g., the ammonia hydrate peritectic temperature in Ceres' pressure conditions is $\sim$176 K (Hogenboom \etal 1997), i.e., similar to Ceres' surface temperature. The excitation process is then related to possible current circulation and loading due to the presence of the fluid reservoir. \tjc{Also, the presence of a deep liquid layer may result in increased dissipation, as recently suggested by Tyler (2008) in the case of outer planet icy satellites.} 
Although the modeling of this process is beyond the scope of this paper, circulation in closed ocean systems and its signature on the rotation is a recent topic of interest to planetary sciences (e.g., Tyler 2008; Noir \etal 2009).  

\subsection{Hydrostatic State}

There are multiple sources of departure from hydrostatic equilibrium at the large scale, starting with the large contrast in temperature between the equator and the poles, of at least 50 K (Fanale and Salvail 1988). Castillo-Rogez and McCord (2010) suggested that Ceres' equatorial temperature may promote the preservation of a regional deep liquid layer while the polar regions would be entirely frozen. The contrast in density between water ice and liquid water saturated in brines may be up to 60\% (e.g., Prieto-Ballesteros and Kargel 2005), which would increase the difference between $A$ and $C$ by about 5\%. Another source of density anomalies are mascons (mass concentrations), for example due to topography anomalies at the silicate core, as inferred for Ganymede from \textit{Galileo} measurements (e.g., Palguta \etal 2009).  Schenk and McKinnon (2008) have suggested in the case of Enceladus that an unrelaxed core is responsible for the departure of the satellite's  shape from hydrostaticity, by $\sim$1.5 km, even if the outer shell of the satellite is likely to have relaxed. 
Topographic features, for example unrelaxed craters are another source of density anomalies. 

The knowledge of the principal axis moment of inertia is key to estimating the departure from hydrostatic equilibrium that is generally assumed in order to interpret degree-two gravity and oblate shape data in terms of interior properties through simple relationships (Equation \ref{eq:J2_I}, Zharkov \etal (1985)). A determination of the mean moment of inertia $I$ from $(A+B+C)/3$ independently from the former equation by using the rotational motion of the body (see Ferrari \etal 1980; Konopliv \etal 2006) would enable the detection of large variations in internal structure. 
Unfortunately, the amplitudes of the nutation and the precessional motion of Ceres are very small. Their measurement requires the tracking of a landmark at the surface of Ceres with an accuracy better than 10 cm, and this for at least six months. Also, since it is unlikely that Ceres' obliquity is fully damped (section \ref{sec:obliq}), we will not be able to rely on the assumption that Ceres is in the Generalized Cassini state as a means to determine the principal axis moments of inertia. Therefore, the comparison of gravity and shape data appears the best approach to infer the presence of non-hydrostatic anomalies (with the Chandler mode, if detected) in the case of Ceres. 

The extent of global relaxation can also be assessed from the comparison of the secular Love numbers inferred independently from the shape $h_s$ and from the degree-two gravity field $k_s$ (e.g. Dermott and Thomas 1988). 

The Love number $k_s$  can be inferred from the degree-two gravity field such as Eq.~\ref{eq:J2_I}. If the object is in hydrostatic equilibrium, then the Love numbers are related by (e.g., Zharkov \etal 1985): 
\begin{equation}
h_s = k_s + 1.
\end{equation}
Departure from this relationship informs on the non-hydrostaticity of Ceres. The \textit{Dawn} Mission is likely to yield the gravity field of Ceres to degree 10 as an outcome of the nominal mission (McCord et al. 2011). 
The ratio of the gravity data to the topography (admittance) is generally used to constrain the degree of isostatic relaxation achieved by geological features (e.g., Simons \etal 1994). For example, Nimmo \etal (2010) interpreted Rhea's degree-three gravity coefficient inferred by Iess \etal (2007) from the \textit{Cassini} Orbiter, as the signature of unrelaxed impact craters. 
Line-of-sight gravity measurements are also most appropriate for detecting lateral variations in density, to be compared against the topography measurements to be inferred from high-resolution imaging.

\section{Conclusion}

We have characterized the main components of Ceres' rotation and quantified them assuming Ceres is differentiated into a rocky core and icy shell. First, our modeling predicts that Ceres' obliquity is not constrained by the dissipative history of the asteroid. However, multiple determinations of Ceres' pole agree that its obliquity lies between 0 and 4 deg. The lower bound suggests that Ceres could be relaxed to a generalized Cassini state. However, considering the very long damping timescale, such a situation is unlikely. \tjc{This uncertainty will be resolved by the {\it{Dawn}} }. In any case, important constraints can also be inferred from combining shape and gravity data. These will yield independent determinations of the secular Love number that will be used to constrain Ceres' hydrostatic state, from which the mean moment of inertia of the asteroid can be inferred.

Then, for the stratified, solid model considered in this study, we established upper bounds on the rigid and non-rigid components of the nutations, polar motion and l.o.d. These appear too small to be inferred from space measurement techniques. Then we identified that a detectable perturbation of Ceres' spin state (wobble) may be the signature of a Chandler mode. This mode would have to be excited by recent large impacts or currents in local liquid reservoirs at depth in order to yield a sizeable signature. This aspect needs to be quantified in detail as it offers the prospect to constrain Ceres' thermal state and geophysical evolution.


\section{Acknowledgements}
The authors wish to thank Jim Williams (JPL) and Richard Gross (JPL) for valuable discussions on the secular orbits of asteroids and excitation of Chandler wobble. The authors are also thankful to the anonymous reviewers who helped improve the quality of this manuscript. 
Part of this work was conducted at the Jet Propulsion Laboratory, California Institute of Technology, under contract to NASA. Government sponsorship acknowledged. All rights reserved. 
\tnr{This research benefits of a financial support from Paris Observatory (2010).}

\end{document}